\documentclass[preprint,amsmath,amssymb,aps,prd,showpacs]{revtex4-1}

\usepackage{epsfig}
\usepackage{dcolumn}
\usepackage{bm}
\usepackage{tikz}
\usepackage{graphicx, graphics, color}
\usepackage{dcolumn}
\usepackage{bm}
\usepackage{hyperref}
\usepackage[mathlines]{lineno}
\usepackage{float}
\usepackage{subfig}

\newcommand{\be}{\begin{equation}}
\newcommand{\ee}{\end{equation}}
\newcommand{\ben}{\begin{eqnarray}}
\newcommand{\een}{\end{eqnarray}}

\begin{document} 

\title{Charge transport in chiral solids as a possible tool in search of dark matter signals}

\author{Marek Rogatko} 
\email{rogat@kft.umcs.lublin.pl}
\author{Karol I. Wysokinski}
\email{karol@tytan.umcs.lublin.pl}
\affiliation{Institute of Physics, 
Maria Curie-Sklodowska University, 
pl.~Marii Curie-Sklodowskiej 1,  20-031 Lublin,  Poland}

\date{\today}

\begin{abstract}
As the existing techniques to directly detect dark matter particles in laboratory experiments continue to produce negative signals, the research community considers novel approaches including those relying on condensed matter systems with non-trivial quantum properties. For example, finding signatures of dark matter in transport characteristics of solids would be an important step on the road to detect this illusive component of the mass of our Universe. As a first step in this direction we have  recently  derived the modified kinetic equation taking into account two coupled $U(1)$-gauge fields, one being the standard Maxwell electromagnetic field and other corresponding to the dark sector. The resulting Boltzmann kinetic equation is modified by the Berry curvature which couples to both visible and dark sector gauge fields. The  linear and nonlinear longitudinal and/or Hall currents in topological matter may signal the  existence of dark particle induced electromagnetic fields. 
The usage of appropriately designed resonant cavity to enhance the electric field may help to reach the goal.
The conservative  estimates show that the technique should have the sensitivity of the order of $10^{-9}$ or better  and  may help to partially fill the gap between XENON1T and haloscope experiments. 
\end{abstract}

\maketitle
\flushbottom

\section{Introduction}
The only evidences of a dark matter~\cite{bertone2018} come from the gravitational observations.  
The signals include galaxy rotation curves, gravitational lensing and the gravitational structures at cosmological scales~\cite{frenk2012}. The search for this elusive, albeit dominant, component of the matter in our Universe by various techniques continues~\cite{gaskins2016} to give negative effects. The direct unique signals of dark matter  in the laboratory experiments  would provide a breakthrough comparable to the first observation of the gravitational waves, which also resulted in the proposal that primordial black holes may in fact constitute the dark matter~\cite{bird2016}. The absence of evidence  for dark-matter particles despite extensive search results in a ``sense of `crisis' in the dark-matter particle community''~\cite{bertone2018a}. 
Observing this state of the affairs the authors of~\cite{bertone2018a}  propose: 
 ``We argue that diversifying the experimental effort and incorporating astronomical surveys and gravitational-wave observations is our best hope of making progress on the dark-matter problem.'' On the other hand,
 this situation provokes the appearance of novel detection techniques and theoretical models of dark matter (DM).

The mass of DM particle is unknown and may span many orders of magnitude \cite{battaglieri2017}. Some novel proposals consider dark matter candidates with huge masses as, e.g., the mentioned~\cite{bird2016} primary black holes or the recently suggested fractionally charged gravitino with the  mass at Planck scale~\cite{meissner2019}. Many of other recent proposals concentrate on the opposite end and consider relatively low masses in the $eV$ range or even much below. The detection schemes of small mass candidates are necessarily very different from the existing techniques. 

Many of the new proposals base on observations of condensed matter phenomena. It is condensed matter in 
which elementary excitations are observed at low energies and may be used as detectors of small mass
dark matter particles. Various novel methods have been recently discussed~\cite{kim2023,dixit2021,liang2018,rogatko2023,hong2022,hochberg2022,hochberg2016,knapen2021,liang2022,chen2022,bloch2022,hochberg2018,budnik2018,knapen2017,knapen2018,bar18} proposing to use both, standard condensed matter systems and quantum materials with topologically non-trivial band structure. The latter
 include {\it inter alia} detecting DM with help of the  superconducting materials~\cite{rog16,rog15a,rog16a,rog17,kic21} or superconducting qubits~\cite{dixit2021}, graphene-based Josephson junctions~\cite{kim2023} or by observing chiral magnetic effect in topologically non-trivial quantum matter~\cite{hong2022,rogatko2023}.

Other detection techniques encompass the search for bosonic dark matter {\it via} absorption in superconductors \cite{hochberg2016}, using superfluid helium \cite{knapen2017}, optical phonons in polar materials \cite{knapen2018},  observations of colour centres production in crystals \cite{budnik2018}, the usage of bulk three-dimensional Dirac semimetals \cite{hochberg2018} or multilayered optical devices \cite{bar18}. The future will tell which of them turns out to be successful.  The usage of various condensed matter systems and detection techniques involving charge carriers, phonons, light and quasiparticles in superconductors to detect light dark matter has been recently reviewed~\cite{kahn2022,mitridate2023}.

The popular candidates for ultralight DM bosonic particles are axions (more generally axion like particles)  and dark photon. They manifest as dark electric (or magnetic) field 
oscillating with frequency $\tilde{\omega}=m_{DP}c^2/\hbar$, where $m_{DP}$ is the dark particle mass, 
$c$ denotes speed of light and $\hbar$ is reduced Planck constant. In this paper we concentrate on the massive dark photon as an important candidate.

Our starting point will be $(3+1)$-dimensional action describing visible sector comprising Maxwell $U(1)$-gauge field and hidden one with
massive $U(1)$-gauge auxiliary field $\tilde{A}_\mu$ (dark photon), interacting with the ordinary Maxwell field by the so-called kinetic mixing term.
The action is provided by
\be
S_{M-dark~ photon} = \int  d^4x  \Big(
- \frac{1}{4}F_{\mu \nu} F^{\mu \nu} - \frac{1}{4} \tilde{F}_{\mu \nu} \tilde{F}^{\mu \nu} - \frac{\alpha}{2}F_{\mu \nu} \tilde{F}^{\mu \nu}- \frac{m_{DP}^2}{2}  \tilde{A}_\mu \tilde{A}^\mu
\Big),
\label{ac dm}
\ee  
where $\alpha$ is  a coupling constant between two sectors,
standing in front of the kinetic mixing term.
 $m_{DP}$ denotes here the mass of dark photon. $F_{\mu,\nu}=\partial_\mu A_\nu -\partial_\nu A_\mu$, 
 and $\tilde{F}_{\mu,\nu}=\partial_\mu \tilde{A}_\nu -\partial_\nu \tilde{A}_\mu$ denote the field strength 
 tensors of the standard and dark photon, respectively.  The coupling $\alpha$ is expected to be $\ll 1$.

 The idea that dark photon can be a candidate for dark matter has been widely exploited on various backgrounds, both theoretically \cite{hol86}-\cite{abe08a} and experimentally \cite{til15}-\cite{an23}.
 The model in question possesses also some possible astrophysical confirmations \cite{jea03}-\cite{bod15}. We have cited only some illustrative examples due to the vast amount of work authorising 
 this blossoming field of researches.

 Dark photons, the ultralight candidates for dark matter, interact with the Standard Model particles via kinetic mixing term. After appropriate rotation and redefinitions,
 one can eliminate kinetic mixing term and arrive at the Lagrangian for dark photon, Maxwell field and electromagnetic current \cite{dphbook}. Thus, on the solid state physics background
 we can say that dark photon interacts with electrons in solids.
More generally,  the scattering of dark particle like axion (in the presence of external magnetic field) or dark photon produces  small electric field oscillating with frequency $\tilde{\omega}$ at the surface of a metal or superconductor~\cite{iwazaki2020,iwazaki2021,kishimoto2022} and this can possibly be detected.  
 
 In the recent paper~\cite{rogatko2023} we considered the chiral magnetic effect induced by dark photon in the Dirac semimetal. Chiral magnetic  effect denotes the current flowing along the applied magnetic field in conducting material with a Berry curvature; in the mentioned paper the current flows along dark magnetic field $\tilde{\textbf{B}}$. In that scenario the very existence of the signal relies 
 on the kinetic coupling, which is a source of the dark magnetic field.  The estimates show that the expected dark magnetic field is small and thus detection of the non-dissipative current will be difficult, but not impossible. In contrast, the work~\cite{hong2022} also proposing measurement of the chiral magnetic effect, assumes application of the external magnetic field and detection of axion-like particles. In that work the current flows along the direction of the external magnetic field $\textbf{B}$.
  
In the present paper, we extend our previous work and consider detection of dark matter  {\it via} measurement 
of anomalous transport in Dirac or Weyl semimetals~\cite{armitage2018,lv2021}, with broken time reversal or inversion symmetry. Dirac or Weyl semimetals are three or two-dimensional condensed matter systems with topologically non-trivial energy spectrum, which is characterized by the existence of Dirac nodes and zero, or very small energy gap between conduction and valence bands. The oscillating electric field induced by dark particles like axions or dark photons in a two-dimensional Weyl or Dirac semimetals, could be identified by observing in a such systems two kinds of electric currents flowing  in mutually perpendicular directions, i.e., the longitudinal and anomalous Hall current. 
The latter appears at linear electric field order if the time reversal symmetry is broken.
On the other hand, at higher order in electric field strength, the anomalous Hall effect appears even when the Hamiltonian is time reversal invariant, but other symmetries like inversion are broken~\cite{sodemann2015}. Two dimensional transition metal dichalcogenides~\cite{wang2018,chang2023} or three-dimensional Weyl semimetals~\cite{armitage2018} could serve as potential candidates to observe the effect and thus as detectors of dark matter fields.  

	The organization of the rest of the paper is following. In Sec. \ref{chiral} we briefly recall the Boltzmann kinetic theory application to chiral systems, like Weyl or Dirac semimetals in the  Maxwell $\textbf{E}$ and dark sector $\tilde{\textbf{E}}$ electric fields. We solve the Boltzmann equation in Sec. \ref{solution}. The currents in the system with two electric fields are calculated in Sec. \ref{currents}. The estimation of the dark particle induced electric field is a subject of Sec. \ref{sec:bfield}, while its possible enhancement in a resonant cavity is discussed in Sec. \ref{sec:enh}. The estimates of the sensitivity of the proposed measurements are presented in Sec. \ref{sec:sensit}.  We end up with the discussion of a few scenarios to identify signal of dark matter in Sec. \ref{scenarios}.

 \section{Boltzmann kinetic equation for systems with Berry curvature and dark matter}
 \label{chiral}
Standard Boltzmann kinetic theory applied to Weyl and Dirac systems requires a modification which takes into account the existence 
of the Berry curvature $\boldsymbol \Omega (\mathbf{k})$, in such materials.  The existence of a non-trivial electron spectrum 
and resulting Berry curvature is known to modify the equations of motion of the electron velocity $\dot{\mathbf{r}}$ and momentum $\dot{\mathbf{k}}$,
as well as, a phase space volume $d\mathbf{r}d\mathbf{k}$~\cite{xiao2010}. 
Strictly speaking, the Boltzmann kinetic equation has its standard form. However, the velocity of a particle is modified by the presence of Berry curvature 
and this induces the modification of the particles' dynamics. 

On the other hand,
Boltzmann equation allows one to find the local distribution function \cite{mahan} $f(r,\textbf{k},t)$ in the presence of external perturbations. In Weyl or 
Dirac semimetals with nodes of different chirality, the distribution may depend on the node in question. Thus, in general we denote the distribution  around the node $c$ 
by $f_c(\mathbf{r},\mathbf{k},t)$. It changes in time according to the relation 
\be
\frac{\partial f_c(\mathbf{r},\mathbf{k},t)}{\partial t}+\mathbf{\dot{r}}_c\cdot\boldsymbol{\nabla}_{\mathbf{r}} f_c(\mathbf{r},\mathbf{k},t)
+\mathbf{\dot{k}_c}\cdot\boldsymbol{\nabla}_{\mathbf{k}} f_c(\mathbf{r},\mathbf{k},t)=I_{coll}[f_c(\mathbf{r},\mathbf{k},t)]. 
\label{boltz-coll} 
\ee
Here $I_{coll}[f_c(\mathbf{r},\mathbf{k},t)]$ is the  collision integral. 

In systems with Berry phase the time dependence of $\textbf{r}$ and $\textbf{k}$ near the node $c$ is found \cite{xiao2010,son2013} and is provided by
\ben \label{quasiclas1}
\mathbf{\dot{r}}_c=\frac{1}{D_c(\mathbf{k})}\left[\mathbf{v}_c+\frac{e}{\hbar}(\mathbf{v}_c\cdot \boldsymbol{\Omega}_c)\mathbf{B}+\frac{e}{\hbar}(\mathbf{E}\times\boldsymbol{\Omega}_c)\right] \\
\hbar\mathbf{\dot{k}}_c=\frac{1}{D_c(\mathbf{k})}\left[-e\mathbf{E}-e \mathbf{v}_c\times\mathbf{B}
-\frac{e^2}{\hbar}(\mathbf{E}\cdot\mathbf{B})\boldsymbol{\Omega}_c\right].
\label{quasiclas2} 
\een
In the above equation $\boldsymbol{\Omega}_c$ denotes the wave vector dependent Berry curvature $\boldsymbol{\Omega}_c(\mathbf{k})$ at node $c$ and $D_c(\mathbf{k})=1+\frac{e}{\hbar}(\mathbf{B}\cdot\boldsymbol{\Omega}_c)$ is the phase space correction factor~\cite{xiao2010}.

Our interest is focused on the
solution of Boltzmann equation which simplifies considerably in homogeneous systems and in the absence of external  magnetic field~\cite{sodemann2015,dantas2021,zhang2021}. 
With such an assumptions of system
homogeneity and absence of magnetic field, the distribution function $f(r,\textbf{k},t)=f(\textbf{k},t)$ is independent on $\textbf{r}$. 
We suppose also the validity of relaxation time approximation. The Boltzmann equation reduces to 
\be
\frac{\partial f(\textbf{k},t)}{\partial t} +\dot{k}^a\frac{\partial f(\textbf{k},t)}{\partial k_a}=-\frac{f(\textbf{k},t)-f_0(\textbf{k},0)}{\tau}.
\label{be-t}
\ee
For simplicity, we neglect here the explicit dependence on $c$ and assume the relaxation time $\tau$ is isotropic and constant.

 In the model with standard electric field $E^a(t)=Re~(E^a_0 e^{i\omega t} )$ and  dark matter electric field $\tilde{E}^a(t)=Re~(\tilde{E}_0^a e^{i\tilde{\omega}t})$, 
 the electron charge (-$e$) changes its momentum as a result of the electric force \cite{rogatko2023}
\be
\hbar \dot{k}^a=-e \Big[ E^a(t)+\frac{\alpha}{2}\tilde{E}^a(t) \Big] 
=-e~Re~\Big( E^a_0 e^{i\omega t}+\frac{\alpha}{2}{\tilde{E}_0^a} e^{i\tilde{\omega}t} \Big),
\label{momentum}
\ee
where $\alpha$ is the coupling constant in the kinetic mixing between two $U(1)$-gauge fields in equation (\ref{ac dm}). It is a source of the change 
of the total electric field acting upon an electron in the background of dark matter sector.

\section{Solution of the Boltzmann equation in the presence of Maxwell and dark matter fields}
\label{solution}
This section is devoted to the solution of the equation (\ref{be-t}). We shall find corrections to the distribution function to  an arbitrary order in the electric fields $\textbf{E}(t)$ and $\tilde{\textbf{E}}(t)$. For a standard Maxwell electric field only, the solutions can be found in Refs.~\cite{sodemann2015,dantas2021,zhang2021}. Even though it is possible to generalize the 
previous work~\cite{dantas2021} and obtain the non-perturbative solution of the equation (\ref{be-t}) we 
 limit the discussion to quadratic order.  To this end, in this section we follow Zhang {\it et al.}~\cite{zhang2021}.

To commence with,  one writes the total distribution function as a sum given by
\be
f(\textbf{k},t)=Re \Big( f_0+f_1+f_2+\cdots f_n+\cdots \Big),  
\label{field-expansion}
\ee
where $f_n$ is assumed to vanish as a power $E_0^n,\tilde{E}_0^n$. Having in mind the relations (\ref{be-t})- (\ref{momentum}) and counting the powers of electric fields, we 
derive the set of equations
\ben \label{eq1}
\tau\partial_t f_0(\textbf{k},t)&=&0, \\
\label{eq2}
\tau \partial_t f_1 +f_1&=&\frac{e\tau}{2\hbar}\left(E_0^{a} e^{i\omega t}+E_0^{*a}e^{-i\omega t}+ \frac{\alpha}{2}\tilde{E}_0^{a} e^{i\tilde{\omega} t}+ \frac{\alpha}{2}\tilde{E}_0^{*a} e^{-i\tilde{\omega} t}\right)\partial_a f_0, \\
\label{eq3}
\tau \partial_t f_2 +f_2&=&\frac{e\tau}{2\hbar}\left(E_0^{a} e^{i\omega t}+E_0^{*a}e^{-i\omega t}+ \frac{\alpha}{2}\tilde{E}_0^{a} e^{i\tilde{\omega} t}+ \frac{\alpha}{2}\tilde{E}_0^{*a} e^{-i\tilde{\omega} t}\right)\partial_a f_1, \\
\cdots , \nonumber \\
\tau \partial_t f_n +f_n&=&\frac{e\tau}{2\hbar}\left(E_0^{a} e^{i\omega t}+E_0^{*a}e^{-i\omega t}+ \frac{\alpha}{2}\tilde{E}_0^{a} e^{i\tilde{\omega} t}+ \frac{\alpha}{2}\tilde{E}_0^{*a} e^{-i\tilde{\omega} t}\right)\partial_a f_{n-1},\\
&etc.& \nonumber
\een
We have used here the shorthand notations $\partial_t=\frac{\partial}{\partial t}$ and $\partial_a=\frac{\partial}{\partial k_a}$. The above equations can be solved order by order. 

Equation (\ref{eq1}) comprises
a statement that the equilibrium distribution function does not depend on time $f_0(\textbf{k},t)=f_0(\textbf{k})$. Further, one assumes that  the initial distribution is  given by the Fermi-Dirac function corresponding to temperature $T$ and chemical potential $\mu$. Namely, we have
\be
f_0(\textbf{k})=\frac{1}{e^{\frac{\varepsilon(\textbf{k})-\mu}{k_BT}}+1}.
\ee
The solution of equation (\ref{eq2}) is simple and and can be written as 
\be
f_1(\textbf{k},t)=Re~\Big( f_1^\omega e^{i\omega t} +f_1^{\tilde{\omega}}e^{i\tilde{\omega}t} \Big),
\ee
where we have denoted
\ben
f_1^\omega&=&\frac{e\tau}{\hbar}\frac{E_0^a~\partial_a f_0}{(1+i\omega \tau)}, \label{f1om} \\
f_1^{\tilde{\omega}}&=& \frac{\alpha}{2}\frac{e\tau}{\hbar}\frac{\tilde{E}_0^a~\partial_a f_0}{(1+i\tilde{\omega} \tau)}.
\label{f1tildeom}
\een 
The second order  term  $f_2(\textbf{k},t)$, for two oscillating fields, introduces various harmonic functions
with frequencies $2\omega,2\tilde{\omega},\omega\pm\tilde{\omega}$. Interestingly, the frequency independent term also contributes at the second order. 
After some algebra one arrives at the relation provided by 
\be
f_2(\textbf{k},t)=Re ~\Big(
f_2^{(0)}+f_2^{(2\omega)}e^{2i\omega t}+f_2^{(2\tilde{\omega})}e^{2i\tilde{\omega}t} 
+ f_2^{(\omega+\tilde{\omega})} e^{i(\omega+\tilde{\omega})t}+f_2^{(\omega-\tilde{\omega})} e^{i(\omega-\tilde{\omega})t} \Big),
\ee
where the frequency dependent coefficients imply
\ben
f_2^{(0)}&=&\left(\frac{e\tau}{\hbar}\right)^2\left[\frac{E_0^{*b}E_0^a~\partial_{ab}f_0}{2(1+i\omega \tau)}
+\left(\frac{\alpha}{2}\right)^2\frac{\tilde{E}_0^{*b}\tilde{E}_0^a~\partial_{ab}f_0}{2(1+i\tilde{\omega}\tau)}\right], \\
f_2^{(2\omega)}&=&\left(\frac{e\tau}{\hbar}\right)^2\frac{E_0^{b}E_0^a~\partial_{ab}f_0}{2(1+i\omega \tau)(1+2i\omega\tau)},\\
f_2^{(2\tilde{\omega})}&=&\alpha^2\left(\frac{e\tau}{2\hbar}\right)^2\frac{\tilde{E}_0^{b}\tilde{E}_0^a~\partial_{ab}f_0}{2(1+i\tilde{\omega} \tau)(1+2i\tilde{\omega}\tau)},\\
f_2^{(\omega+\tilde{\omega})}&=&\alpha\left(\frac{e\tau}{2\hbar}\right)^2\left[\frac{E_0^{b}\tilde{E}_0^a~\partial_{ab}f_0}{(1+i\tilde{\omega} \tau)}
+\frac{\tilde{E}_0^{b}{E}_0^a~\partial_{ab}f_0}{(1+i{\omega} \tau)}\right]\frac{1}{[1+i(\omega+\tilde{\omega})\tau]},\\
f_2^{(\omega-\tilde{\omega})}&=&\alpha\left(\frac{e\tau}{2\hbar}\right)^2\left[\frac{E_0^{b}\tilde{E}_0^{*a}~\partial_{ab}f_0}{(1-i\tilde{\omega} \tau)}
+\frac{\tilde{E}_0^{*b}{E}_0^a~\partial_{ab}f_0}{(1+i{\omega} \tau)}\right]\frac{1}{[1+i(\omega-\tilde{\omega})\tau]}.
\een 
The higher order corrections can be obtained in a similar way, if needed.  In the next section we use the above formulae to calculate the ac currents in chiral system.

\section{Currents in chiral systems affected by the dark field}
\label{currents}
We are mainly interested in detecting dark matter fields which may arise in the rare scattering processes  of dark particles with ordinary matter particles. For this purpose, we analyse the currents flowing in general systems, i.e., those with and without Berry curvature and only later we discuss possible detection scenarios. For chiral systems the velocity, the distribution function and the Berry curvature,
depend on the node index $c$. 
However, in the following formulae we neglect this dependence and restrict our attention to the discussion of the currents to the second order in the field strength. To find the total current one has to take all nodes existing in a given material.

In the kinetic equation approach the current flowing in d-dimensional system is calculated from the relation 
\be
j^a(t)=-e\int \frac{d^dk}{(2\pi)^d} \dot{r}^a f(\textbf{k},t),
\label{curr}
\ee
and can be found to an arbitrary order in electric field. However, in a chiral system influenced by dark matter sector, 
the velocity $\dot{\mathbf{r}}$ is modified by the Berry curvature and dark electric field~\cite{rogatko2023}
\be
\dot{r}^a=v^a(\textbf{k})-\frac{e}{\hbar}\epsilon^{abc}\Omega_b (\textbf{k})
\Big( E_c(t)+\frac{\alpha}{2}\tilde{E}_c(t) \Big).
\label{an-velocity}
\ee
In the linear (in electric fields) order, the longitudinal currents 
flow whatever the symmetry of the crystal, however the existence of Hall effect requires breaking the time reversal symmetry. In the non-linear regime the Hall 
current can be measured even in time reversal symmetric systems. It appears at second~\cite{sodemann2015} 
or higher~\cite{zhang2021,zhuang2023,zhang2023,xiang2023} orders in the electric field.

We calculate the current from the relation (\ref{curr}), with the usage of the equation (\ref{an-velocity}), 
first expanding the distribution function with respect to powers of electric fields, as given  in (\ref{field-expansion}) and (\ref{momentum}). 
To the lowest order, i.e., using $f(\textbf{k},t)=Re( f_0)$, one obtains
\be
j_0^a(t)=-e\int \frac{d^dk}{(2\pi)^d} \left[v^a(\textbf{k})-\frac{e}{\hbar}\epsilon^{abc}\Omega_b~
Re \Big(
E_c(t)+\frac{\alpha}{2}\tilde{E}_c(t)\Big) \right]f_0(\textbf{k}).
\ee
This can be rewritten as follows:
\be
j_0^a(t)=Re~\Big(j_0^a +j_0^{a (\omega)} e^{i\omega t}+j_0^{a (\tilde{\omega})} e^{i\tilde{\omega} t}\Big),
\ee
where we have denoted
\ben
j_0^{a(0)}&=&-e\int \frac{d^dk}{(2\pi)^d} v^a(\textbf{k}) f_0(\textbf{k}), \\
j_0^{a(\omega)}&=&\frac{e^2}{\hbar}\int \frac{d^dk}{(2\pi)^d} \epsilon^{abc}\Omega_b E_{(0)c} f_0(\textbf{k}), \label{ahe} \\
j_0^{a(\tilde{\omega})}&=&\frac{\alpha}{2}\frac{e^2}{\hbar}\int \frac{d^dk}{(2\pi)^d} \epsilon^{abc}\Omega_b \tilde{E}_{(0)c} f_0(\textbf{k}). \label{ahe-dark}
\een 
The total current related to the unperturbed distribution function $f_0$ consists of three terms. 
The first of them is proportional to the velocity $v^\alpha(\textbf{k})$ and is independent of the electric field. 
This rectified current $j_0^{a(0)}$  is purely classical. 
Its existence does not require Berry curvature. However, it vanishes due to the fact that
velocity is an odd function of momentum $v^a(\textbf{k})=-v^a(-\textbf{k})$. 
This is correct, as  there can not be a current flow in the equilibrium. 

On the other hand, the anomalous velocity  induced by the Berry curvature (second term in Eq. (\ref{an-velocity}))  also gives  the currents 
proportional to the unperturbed distribution function and to electric field. To the first order in the field, the current possesses two harmonic contributions 
bounded with the standard and dark field of frequency $\omega$ and $\tilde{\omega}$, respectively. It is this second contribution, that appearance
signals the dark sector effect. This is true, even in the experimental setup with $\omega=0$, i.e., for the static standard electric field $E^a(t)=E_0^a$. 

For $f(\textbf{k},t)$ linear in both fields ($f_1(\textbf{k},t)$), one finds the current, which contains terms linear and quadratic in electric fields. 
The linear terms  result from the velocity $\textbf{v}$, while quadratic ones are due to $ \textbf{E} \times \mathbf{\Omega} $ term in relation (\ref{an-velocity}). 
Thus, we obtain the total time dependent current consisting of various components, each oscillating at a specific frequency 
\ben
j_1^a(t) &=& Re ~\Big(
j_1^{a (0)}+j_1^{a (\omega )}e^{i\omega t} +j_1^{a (\tilde{\omega})} e^{i\tilde{\omega}t} 
+  j_1^{a (2\omega)}e^{2i\omega t} \\ \nonumber
&+& j_1^{a (2\tilde{\omega})} e^{2 i\tilde{\omega}t}+
j_1^{a (\omega+\tilde{\omega})} e^{ i(\omega+\tilde{\omega})t}+j_1^{a (\omega-\tilde{\omega})} e^{ i(\omega-\tilde{\omega})t} \Big),
\een
 where the various coefficients are given by the following formulae:
\ben
j_1^{a (0)}&=&\frac{e^2}{2\hbar}\int \frac{d^dk}{(2\pi)^d} \epsilon^{abc}\Omega_b \left[E_{(0)c} f_1^{*(\omega)} +\frac{\alpha}{2}\tilde{E}_{(0)c} f_1^{*(\tilde{\omega})} \right], \label{rect1}\\
j_1^{a (\omega)}&=&-e\int \frac{d^dk}{(2\pi)^d} v^a(\textbf{k})f_1^{(\omega)}, \\
j_1^{a (\tilde{\omega})}&=&-e\int \frac{d^dk}{(2\pi)^d} v^a(\textbf{k})f_1^{(\tilde{\omega})}, \label{1cur-dark} \\
j_1^{a (2\omega)}&=&\frac{e^2}{2\hbar}\int \frac{d^dk}{(2\pi)^d} \epsilon^{abc}\Omega_b E_{(0)c} f_1^{(\omega)},\label{lin-curr} \\
j_1^{a (2\tilde{\omega})}&=&\frac{\alpha}{2}\frac{e^2}{2\hbar}\int \frac{d^dk}{(2\pi)^d} \epsilon^{abc}\Omega_b \tilde{E}_{(0)c} f_1^{(\tilde{\omega})}, \label{j1dark} \\
j_1^{a (\omega+\tilde{\omega})}&=&\frac{e^2}{2\hbar}\int \frac{d^dk}{(2\pi)^d} \epsilon^{abc}\Omega_b \left[E_{(0)c} f_1^{(\tilde{\omega})} +\frac{\alpha}{2}\tilde{E}_{(0)c} f_1^{(\omega)} \right], \label{jom-plus-tom}\\
j_1^{a (\omega-\tilde{\omega})}&=&\frac{e^2}{2\hbar}\int \frac{d^dk}{(2\pi)^d} \epsilon^{abc}\Omega_b \left[E_{(0)c} f_1^{*(\tilde{\omega})} +\frac{\alpha}{2}\tilde{E}_{(0)c}^* f_1^{(\omega)} \right],
\label{jom-min-tom}
\een 
with $f_1^{(\omega)}$ and $f_1^{(\tilde{\omega})}$ given by the relations
(\ref{f1om}) and (\ref{f1tildeom}). Note, that the latter function contains the factor $\frac{\alpha}{2}$, which makes both coefficients (\ref{jom-plus-tom}) and (\ref{jom-min-tom}) to be 
linear in the coupling constant $\alpha$.

\section{Estimation of the dark photon induced electric field}
\label{sec:bfield}
The very recent measurements of the local {dark matter} densities conducted by LAMOST DR5 and Gaia DR2 experiments \cite{guo20,loe22}, reveal that our Galactic 
disc is immersed in {dark matter} halo with a characteristic mass density $\rho_{DM}=0.5 GeV/cm^3 $. 
In addition, the latest 
observations of the gravitational lensing seem to indicate that the model of wave-like dark matter 
better fits the observed lensing features ('brightnesses and positions of multiply-lensed images'), then the model of particle-like dark matter~\cite{amruth2023}. 

The {dark matter} density is  expected to be bounded with the amplitude of the dark photon, and time and space components of its vector potentials $\tilde{{A^0}}$ and $\tilde{\textbf{A}}$, 
which are subject to time and space dependences~\cite{kishimoto2022}
\ben
\tilde{A}^0(\textbf{r},t)&=&-\tilde{\textbf{v}}\cdot\tilde{\textbf{A}}_c\cos(m_{DP}t-m_{DP}\textbf{v}\cdot\textbf{r}),\\
\tilde{\textbf{A}}(\textbf{r},t)&=&\tilde{\textbf{A}}_c\cos(m_{DP}t-m_{DP}\textbf{v}\cdot\textbf{r}).
\een
Furthermore,
the time component of the dark photon field is suppressed by the velocity of dark sector around the Earth $|\textbf{v}|\approx 10^{-3}c$. The field 
oscillates with frequency $m_{DP}$ and thus provides a window for its detection by frequency dependent measurements. The amplitude $\tilde{\textbf{A}}_c$ 
is related to the density of the dark matter $\rho_{DM}$, depending on {\it dark photon} mass \cite{beringer2012,iwazaki2020,iwazaki2021,kishimoto2022}
\be
\rho_{DM}=\frac{1}{2}m^2_{DP}\tilde{\textbf{A}}^2_c.
\label{rhoDM}
\ee
The dark photon induced electric field $\textbf{E}$ at the surface of the metal with conductivity $\sigma$ implies \cite{kishimoto2022} 
\be
{ \textbf{ E}}=\alpha \sqrt{\frac{m_{DP}}{2\sigma}} \tilde{\textbf{A}}_c[\cos(m_{DP}t)-\sin(m_{DP}t)]=\alpha \sqrt{\frac{m_{DP}}{\sigma}} \tilde{\textbf{A}}_c\cos(m_{DP}t+\frac{\pi}{4}) .
\label{elfield-DM}
\ee
The dark photon induced electric field oscillates with frequency $\tilde{\omega}=\frac{m_{DP}c^2}{\hbar}$ 
and is phase shifted by $\pi/4$, in comparison to the behaviour assumed in Sec. \ref{chiral}. 

 Combining (\ref{rhoDM}) and (\ref{elfield-DM}) one finds the magnitude of the dark matter induced electric field amplitude 
\be
\tilde{\textbf{E}}_0=  c\sqrt{2\mu_0\rho_{DM}},
\ee
in standard units. Using dark matter density 0.5GeV$/cm^3$ we get $\tilde{\textbf{E}}_0\approx 130\frac{V}{m}$. This field oscillates at frequency in GHz regime, induces in vacuum the electric field, oscillating with the same frequency and small magnitude. Namely one has
\be
\textbf{E}_0=\alpha \tilde{\textbf{E}}_0,
\label{ind-el-field}
\ee
 as discussed earlier.

\section{Enhancing the induced field}\label{sec:enh}
The induced electric field is scaled down by the kinetic mixing term $\propto$ coupling. Due to the fact that $\alpha\ll 1$ it would be of great importance to amplify its value.  One possibility is to implement the cavity resonators. In such devices the electric field is reflected from two parallel mirrors distance $L$ apart from  each other. The wave moving towards one the mirrors is reflected back 
changing the sign of the wave-vector and the sign of the amplitude. The appropriate boundary conditions require the vanishing of the field component parallel to the mirror. Depending on the details of the setup the electric field inside the resonator may be further increased. 

The reflection process is always related to some losses of the signal. However, for high quality factor $Q$ of the resonator the reflection coefficient $r$ is close to unity. This is because~\cite{gue2023}
\be
r=1-\frac{\pi}{2Q}.
\ee
In typical resonators the standing waves form and this results in a very inhomogeneous electromagnetic fields. The field inhomogeneity makes standard resonators not applicable for the purpose of transport measurements and more complicated systems are needed. This can be achieved by supplementing the resonators with tuners (metallic reflectors) which change the boundary conditions and may give more homogeneous fields, at the expense that not all frequencies can be effectively used. With large quality factors one may get a resonator of arbitrary size and working frequency. The details depend on the size, shape and construction of tuners. Smaller resonators allow for larger usable frequencies. Such devices are usually called electromagnetic reverberation chambers~\cite{hill2009}. It is a complicated task to calculate the electric field enhancement in a resonator of arbitrary shape. 

However, the order of magnitude estimation may be obtained by considering the simplest possible device consisting of two parallel metallic mirrors similar to that considered in~\cite{gue2023}. It has been shown that both the external electric field $\textbf{E}$ and the dark matter introduced field $\tilde{\textbf{E}}$ get enhanced. The idea  is as follows. Consider 
a resonator in a form of parallelepiped with two mirrors characterized by
the reflection coefficient $r$. The electric field propagating with a wave vector $\textbf{k}$ and frequency $\omega$ as a plane wave $\textbf{E}=Re[\textbf{E}_0e^{i(\textbf{k}\cdot\textbf{r}-\omega t)}]$, with the amplitude $\textbf{E}_0$, is reflected back on the mirror with probability $r$ changing the direction and wave vector to opposite one. During its travel from one mirror to another it also gets additional phase $e^{i{kL}}$, where $k$ is component of $\textbf{k}$ along the direction between the mirrors. After many such reflections, the resulting total field $\textbf{E}_{tot}(t)$ at $x=0$, which is taken to be in the middle of the cavity, is found to be \cite{gue2023}
\be
\textbf{E}_{tot}(t)= Re\left[\textbf{E}_0\frac{e^{i\frac{kL}{2}}}{1+re^{ikL}}e^{-i\omega t}\right].
\ee    
Comparing this relation with our definition of the field amplitude 
given before equation (\ref{momentum}), one notices that the factor 
\be
\textbf{E}_{0tot}=\textbf{E}_0\frac{e^{i\frac{kL}{2}}}{1+re^{ikL}},
\label{E-tot}
\ee
 plays a role of $\textbf{E}^*_0$, used previously.  Thus for an experiment utilising cavity one replaces everywhere in the previous formulae $\textbf{E}_0$ by $\textbf{E}_{0tot}^*$. 

As far as the dark matter field is concerned, we have to notice that according to the observations, it exists everywhere in the space around the Earth, moving with relative velocity $v\approx 10^{-3}c$ and is considered as a coherent over typical experimental distances. This leads to the conclusion that it produces the electric field $\alpha \tilde{\textbf{E}}$ in both mirrors. However, only the component perpendicular to the mirrors, denoted $\tilde{\textbf{E}}_\perp$,  propagates and in effect only this component is enhanced. Adding the constant in space component one finds that the total dark matter field inside the cavity at $x=0$, is given by \cite{gue2023}
\be
\tilde{\textbf{E}}_{tot}(t)=\left[\tilde{\textbf{E}}_0+2\tilde{\textbf{E}}_{0\perp}\frac{e^{i\frac{kL}{2}}}{1+re^{ikL}}\right]e^{-i\tilde{\omega}t}.
\ee
Once again we notice that the role of a complex amplitude of the dark matter field  $\tilde{\textbf{E}}^*_0$, used previously, is played  by the following relation:
\be
\tilde{\textbf{E}}_{0tot}=\left[\tilde{\textbf{E}}_0+2\tilde{\textbf{E}}_{0\perp}\frac{e^{i\frac{\tilde{k}L}{2}}}{1+re^{i\tilde{k}L}}\right],
\label{E-tot-dm}
\ee  
where the frequency and the wave vector of dark matter massive field  in the laboratory frame is calculated (for arbitrary boost direction) from
\ben
\frac{\tilde{\omega}{'} }{c}=\gamma(\frac{\omega}{c}-\textbf{v}\cdot\textbf{k})\approx\frac{\omega}{c},\\
\tilde{\textbf{k}}{'}=\tilde{\textbf{k}}+\frac{1}{v^2}(\gamma-1)(\tilde{\textbf{k}}\cdot\textbf{v})\textbf{v}-\gamma\frac{\omega\textbf{v}}{c^2},
\een
where $\gamma=1/\sqrt{1-\frac{v^2}{c^2}}$.
The approximate equality in the first equation is motivated by 
the fact that
$\gamma\approx 1$. Similarly, this shows that $\tilde{\textbf{k}}{'} \approx -\frac{\omega \textbf{v}}{c^2}$, for the  dark photon mode  of interest, $\tilde{\textbf{k}}=0$. The wave vector $\tilde{\textbf{k}}'$ can be neglected for macroscopic resonators, when $\tilde{k}'L\ll2\pi$. The massless photon moves with $k=\frac{\omega}{c}$.     

To proceed, we rewrite (\ref{E-tot}) and (\ref{E-tot-dm}) in terms of photon frequencies as 
\ben
\textbf{E}_{0tot}=\textbf{E}_0\frac{e^{i\frac{\omega L}{2c}}}{1+re^{i\frac{\omega L}{c}}}=\textbf{E}_0 B(\omega), \label{enh-b}\\
\tilde{\textbf{E}}_{0tot}
=\tilde{\textbf{E}}_0\left[1+2\beta\frac{e^{i\frac{\tilde{\omega}L}{2c}}}{1+re^{i\frac{\tilde{\omega}L}{c}}}\right]=\tilde{\textbf{E}}_0 A(\tilde{\omega}),
\label{enh-a}
\een 
and the parameter $\beta=1/\sqrt{3}$ characterises the average value of the projection of dark photon field on the unit vector $\hat{\mathbf{n}}$ perpendicular to the mirror
\cite{gue2023}. The frequency dependent factors $A(\tilde{\omega})$ and $B(\omega)$ provide the appropriate complex amplification amplitudes.

\begin{figure}[h]
\includegraphics[width=0.85\linewidth]{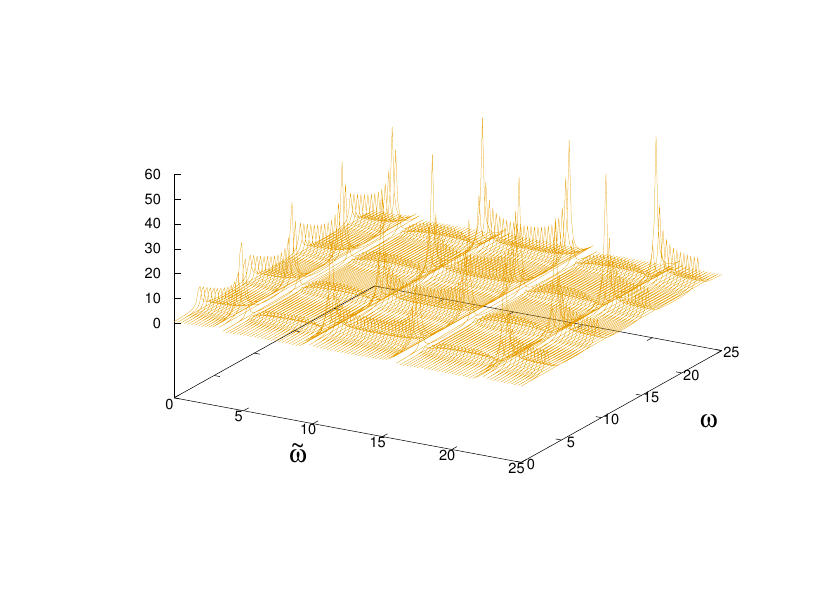}
\caption{(color online) The dependence of the modulus of amplification factor $W(\tilde{\omega},\omega)=A(\tilde{\omega})B^*(\omega)$ 
of the $\textbf{j}_1^{\omega-\tilde{\omega}}$ current defined in Eq. (\ref{jom-min-tom}). We measure frequencies in unit of $c/L$ and assume the modest value of the reflection coefficient $r=0.95$.  }
\label{fig:rys1} 
\end{figure}

According to the previous discussion the transport experiments aimed at the discovery of dark matter will concentrate on the currents related to dark photon induced electric field. These include the anomalous Hall current (\ref{ahe-dark}), which existence does not require external field. Its value is enhanced by the  
amplitude $A(\tilde{\omega})$. Similarly, the longitudinal current (\ref{1cur-dark}), which also does not require external field is amplified by the same factor. The remaining two currents which are of linear order in $\alpha$, namely 
relations (\ref{jom-plus-tom}) and (\ref{jom-min-tom}) can be rewritten in terms of bar amplitudes and enhancement factors. One finds that the current depending on the sum of frequencies is enhanced by $A^*(\tilde{\omega})B^*(\omega)$, while that corresponding to the difference of frequencies is enhanced by the complex amplitude given by $A(\tilde{\omega})B^*(\omega)$. These enhancement factors are complex and their maximal values depend on the reflection coefficient $r$. For cavities with large $Q$ values $r$ is very close to 1 and the amplification factor may, for particular frequencies, attain very large values.

For the 
illustration purposes we use $r=0.95$ and calculate the enhancement as a function of frequency $\omega$ and $\tilde{\omega}$, in units of $c/L$. 
In the figure we plot the absolute value of the amplification factor $W(\tilde{\omega},\omega)=A(\tilde{\omega})B^*(\omega)$ which enters the current $\textbf{j}_1^{\omega-\tilde{\omega}}$ defined by 
the relation (\ref{jom-min-tom}). The  amplification factor is large and gets larger with the increase of the $r$ value. For $r=0.999$ it reaches the value as large as a few thousands.

\section{Sensitivity analysis}
 \label{sec:sensit}
As discussed in previous sections, at the linear order in the coupling $\alpha$, the signals from dark photon of frequency $\tilde{\omega}$ which we expect to detect by using external high frequency $\omega$ radiation may appear at the set $\{\tilde{\omega},\omega+\tilde{\omega}, \omega-\tilde{\omega}\}$ of frequencies. Signal at frequency $\tilde{\omega}$ appears at the linear order of the field $\mathbf{\tilde{E}}$, while the other at the second order of fields. As visible from the
equations
 (\ref{lin-curr},\ref{jom-plus-tom},\ref{jom-min-tom}) the signals are proportional to $\mathbf{\tilde{E}}$ in the linear case and the combinations of the fields $\mathbf{\tilde{E}}\mathbf{E}$ enter the high frequency anomalous non-linear Hall signals. It has to be mentioned that high frequency transport measurements of the longitudinal and Hall currents can utilise various techniques. The analysis of the projected experimental sensitivity may depend on the particular measurement technique. Here we concentrate on one source of experimental uncertainty which is  the fluctuations in the amplitude of the field.  The estimate of the limit on the coupling $\alpha$ due to this source are analogous to those in Ref.~\cite{gue2023} and we shall follow their analysis. 

The anomalous Hall currents we are interested in are of linear order in $\alpha$ and proportional to the products of two fields. For the measurements in the cavity one has to note that the cavity enhances both the field and its fluctuations. The latter are related to the power fed into the cavity and to the power fluctuations.  Modelling the injected field amplitude by stochastic component means that the field reads $\mathbf{E}(\mathbf{r},t)= Re\left[\mathbf{E}_0e^{i(\mathbf{k}\mathbf{r}-\omega t)}+\frac{\Delta \mathbf{E}_0}{2}(e^{i(\mathbf{(k-k_0)}\mathbf{r}-(\omega-\omega_0) t)}+e^{i(\mathbf{(k+k_0)}\mathbf{r}-(\omega+\omega_0) t)})\right]$. Here, $k_0,\omega_0$ is the wave vector and frequency of the amplitude modulations. In principle, all involved phases can be shifted by arbitrary constants, but we neglect here such phase shifts for simplicity. 
In analogy to the calculations in Section (\ref{sec:enh}) we propagate all  components of the field with fluctuating amplitude and get at $\mathbf{r}=0$ 
\be
\mathbf{E}_{tot}(t)=Re\left[\mathbf{E}_0 B(\omega)e^{-i(\omega t +\phi)}+\frac{\Delta \mathbf{E}_0}{2} B(\omega_{-})e^{-i(\omega_{-} t +\phi_{-})}+\frac{\Delta \mathbf{E}_0}{2} B(\omega_{+})e^{-i(\omega_{+} t +\phi_{+})}\right].
\ee
We reintroduced arbitrary phases $\phi$ and $\phi_0$ and denoted $\phi_{\pm}=\phi\pm\phi_0$, $\omega_{\pm}=\omega\pm\omega_0$. The function $B(\omega)=\frac{e^{i\frac{\omega L}{2c}}}{1+re^{i\frac{\omega L}{c}}}$  is the same enhancement factor as in the relation (\ref{enh-b}).

\begin{figure}[h]
\includegraphics[width=0.85\linewidth]{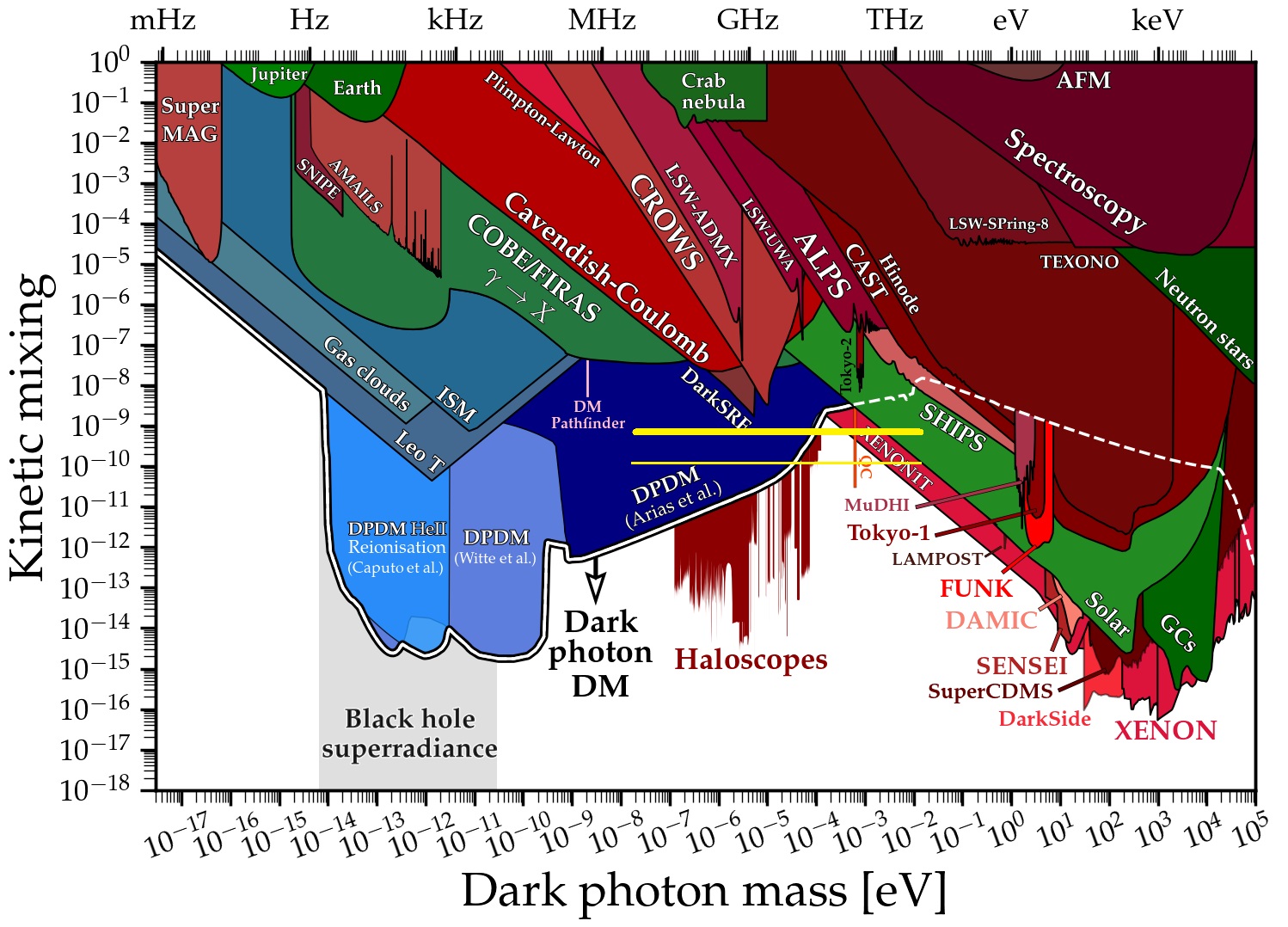}
\caption{(color online) The expected constrained of the discussed technique on the kinetic mixing $\alpha$ is marked on the standard diagram~\cite{sensitivity}  by the yellow solid lines. Thick line is for the conservative and thin one for the optimistic sensitivity estimate  as discussed in the text. In both cases the proposed technique allows for partial filling of the gap between haloscopes' and XENON1T's measurements.   }
\label{fig:rys2} 
\end{figure}

To be on the save side one should take into account the enhanced fields at the minimum and enhanced noise at the maximal values. This leads to the conservative estimation. The order of the magnitude calculation of the accuracy, using the parameters relevant for our proposal of the same value as in~\cite{gue2023}, leads to 
a rough sensitivity estimation of the order of $\alpha\approx 10^{-10}\div 10^{-9}$.  Importantly, even with the conservative estimate the technique may be of importance, as it fills the gap between the existing measurements by XENON1T and haloscope experiments as visible from the Fig. (\ref{fig:rys2}). In the figure the thick yellow horizontal line shows the limit of the kinetic mixing corresponding to our sensitivity estimate $=10^{-9}$. The expected increase of sensitivity to $10^{-10}$ due to correlation between two signals at two frequencies  further closes the existing gap. This is represented by the thin yellow line in the figure. 

We now provide a discussion of the experimental setup and the mass range which can be accessed by the present technique. The measurements of ac longitudinal and anomalous Hall currents are among standard tools in the condensed matter field. The limitation on the upper limit of the  dark photon mass is related to its expected de Broglie wave-length $\lambda_{DP}=\frac{h}{m_{DP}v_{DP}}$, which should be greater than the typical size of the measured sample $L_s$. Using $L_s\approx 1 mm \div 1 cm$, $v_{DP}\approx 10^{-3}c$ one gets the upper mass limit $\approx 6\div 60 meV$. Knowing the dark matter density $\rho$ and assuming that it is composed of dark photons of the above mass only we estimate that there exist about $10^{11}\div 10^{10}$ of them in each cubic $cm$ of space around Earth. 

The limit on the lower mass range depends on the particular detection scheme. We  discuss here  the most promising scheme requiring simultaneous measurements of non-linear anomalous Hall currents at two frequencies $\omega\pm\tilde{\omega}$, distance $2\tilde{\omega}$ apart. If the frequency $\tilde{\omega}$ is too small the total frequency shift $2\tilde{\omega}$ may not exceed the width of the experimental signal peak making the frequency shift unobservable. Even though the precise numbers do depend on experimental setup we expect that other unexplored mass range, namely  $m_{DP}<10^{-14}$ eV ($c.f.$ wide unexplored range of masses in Fig. (\ref{fig:rys2})), is not accessible in the current scheme.

By definition the observed currents do not depend on the mass of the sample or the acquisition time.  Moreover, the sensitivity estimate does not depend on the mass/frequency range and  is thus valid from the lowest (in actual experimental practice of order of hundred Hz) up to THz frequencies.

Additionally to the above estimates on the sensitivity we propose another piece of information related to the issue and a particular technique of measuring the anomalous Hall current. One way to learn about the high frequency anomalous Hall conductance of materials is to measure the change of the angle between the beams of linearly polarised light incident and reflected from the surface of material displaying the Hall effect~\cite{shimano2002}. This technique known as Kerr effect has been widely utilised in measuring Hall effect in superconductors with time reversal breaking order parameter~\cite{kapitulnik2009} without external magnetic field. To measure the tiny effects in Sr$_2$RuO$_4$~\cite{kiw2012}  and other superconductors special instrument was built~\cite{xia2006} reaching the accuracy of Kerr angle measurements as small as a few $nrad$ at energy of light quanta  $=0.8 eV$. Combining this with recent calculations~\cite{sonowal2019} which show that in Weyl semimetals the Kerr angle is of order of $0.1 rad$ one gets the realistic estimate of sensitivity for the existing apparatus and typical materials of the order of $10^{-8}$. 
We propose to measure three different signals, so the statistical correlations among them can be utilised to  identify the events even if the individual measurements are less precise.  


\section{Discussion of detection scenarios and conclusions}
\label{scenarios}
We consider massive dark photon as a source of the electric field which couples to the standard one by kinetic mixing characterised by the coupling $\alpha$. With both, Maxwell and dark photon fields of respective frequencies $\omega$ and $\tilde{\omega}$ taken into account, we  have solved the Boltzmann equation up to the first order in the fields, 
although, the proposed scheme is valid to all orders. Moreover, the closed non-perturbative  solution (valid to arbitrary order of the electric fields) can also be obtained by straightforward generalization of the recent paper~\cite{dantas2021}. 
%
The alternating longitudinal and anomalous Hall currents induced by the sum of the external electric field and the field due to the dark photon  are calculated up to the second order in the fields and up to the linear order in the coupling $\alpha$.  There are three contributions of order of $\alpha^1$. The conservative estimations show the sensitivity which sets the limit about $10^{-9}\div 10^{-10}$.  The proposed experiments may fill the gap between the XENON1T and haloscope measurements, as visible from the Fig. (\ref{fig:rys2}). The existing instrumentation should potentially allow detection of the effect in Weyl semimetals. 

Various detecting scenarios can be thought about in the context of the present proposal. 
Nevertheless, one has to recall that the $\alpha$-coupling constant in the 
kinetic mixing term is expected to be small, $\alpha \ll 1$, so the terms of order $\alpha^2$ can be probably neglected as unmeasurable, even with application of resonant electromagnetic cavities. This is true for the second term of the rectified current (\ref{rect1}) and for the current (\ref{j1dark}) which is also proportional to $\alpha^2 \tilde{\textbf{E}}^2_{(0)}$.

It is important to have a possibility to measure all currents which are of the first order in $\alpha$. These are of the longitudinal and the anomalous Hall character and show up at various field orders and frequencies as discussed earlier. Thus the material to be used should have a band structure allowing for the Berry curvature $\boldsymbol{\Omega(\textbf{k})}$ at some wave vector(s) $\textbf{k}$. Only in such systems with two (or more) wave vectors not related by time reversal symmetry there is a possibility to observe anomalous Hall effect in the linear order, i.e., $\propto \tilde{\textbf{E}}$. The appearance of the nonlinear anomalous Hall effect does not require breaking of time reversal symmetry. 

The expectation is that some of the two dimensional transition metal dichalcogenides or other materials~\cite{pronin2021} with non-vanishing Berry curvature could serve the purpose. These materials are relatively clean and their two dimensional character make the high frequency transport measurements  easier than their three dimensional counterparts.  The expectation that the two dimensional system may in some cases be better probes comes from the observation that, e.g., in graphene a very large non-linear Kerr effect~\cite{zhang2012} has been measured. However, not the dimensionality itself, but the distance $\Delta k_W$ between the Weyl nodes  seems to be the most important parameter, as the recent calculations suggest~\cite{sonowal2019}, because the Kerr signal is directly proportional to it.

We mention here  the most obvious scenarios, leaving more detailed analysis to the future work. One can imagine (i) measuring the  current (\ref{ahe-dark}), which is the {\it ac} analogue of the anomalous Hall~\cite{haldane2004} current in the topological material with broken time reversal symmetry. In a similar manner (ii) the longitudinal current given by (\ref{1cur-dark}) is linear in $\tilde{\textbf{E}}_{(0)}$. This current can also be spotted in a standard metallic system without Berry curvature. Both these currents appear  when  the dark particle occasionally collides with the studied material and produces oscillating electric field $\propto \alpha \tilde{\textbf{E}}(t)$ in the absence of external standard electric field  $\textbf{E}=0$. 

Other scenario (iii) relies on the application of a strong ac standard electric field $\textbf{E}$ of frequency $\omega$ and studying longitudinal and Hall currents at frequencies $\omega\pm\tilde{\omega}$. This is potentially very precise measurement which, however, requires the knowledge of frequency $\tilde{\mathbf{\omega}}$~\cite{zong2023}. Unfortunately, the dark field frequency is unknown and thus the apparatus and measurement details have to be tuned for each expected mass, in a way very much similar  as in other existing or proposed experiments. 
However, the value of the applied frequency $\omega$ may allow to entangle the signals as resulting from the  matter as they appear symmetrically with respect to applied frequency $\omega\pm\tilde{\omega}$. 
 
Surely, the simultaneous observation of all three currents is potentially the most desired scenario.  The rough estimations suggest that the technique sensitivity could be better than $10^{-10}$, especially if the temporal correlations among three different signals will be utilised. 

 Our proposal  supplements other techniques based on the condensed matter systems. The important aspect of the present proposal is the possibility to measure essentially  simultaneously a few currents using  topological quantum materials (with Berry curvature). One expects the existence of three signals  with one longitudinal and two ac anomalous Hall currents. They appear at a few frequencies, e.g., $\tilde{\omega},\omega\pm\tilde{\omega}$, what provides additional control on the effect and  
an additional experimental knob, which can be effectively used to identify the mass of the dark photon. 

 In summary, we have proposed to search for the dark matter induced electric fields by means of the transport currents in chiral Weyl or Dirac semimetals. The discussed here detection scheme of the dark matter is based on the observations of ac longitudinal and Hall currents.

\acknowledgments
K.I. W. and M. R. were partially supported by Grant No. 2022/45/B/ST2/00013 of the National Science Center, Poland.




\end{document}